\documentclass[aps,prl,twocolumn,groupedaddress,showpacs,amssymb,amsmath]{revtex4}
\usepackage{dcolumn}
\usepackage{graphicx}

\begin{document}


\title{Magnetic Correlations at Graphene Edges}

\author{Oleg V. Yazyev}
\email[Electronic address: ]{oleg.yazyev@epfl.ch}
\affiliation{Ecole Polytechnique F\'ed\'erale de Lausanne (EPFL),
Institute of Theoretical Physics, CH-1015 Lausanne, Switzerland}
\affiliation{Institut Romand de Recherche Num\'erique en Physique
des Mat\'eriaux (IRRMA), CH-1015 Lausanne, Switzerland}
\author{M. I. Katsnelson}
\affiliation{Institute for Molecules and Materials, Radboud
University of Nijmegen, Toernooiveld 1, 6525 ED Nijmegen, The
Netherlands}

\date{\today}

\pacs{75.75.+a, 
75.40.Cx, 
81.05.Uw, 
85.75.-d} 


\begin{abstract}

Magnetic zigzag edges of graphene are considered as a basis for
novel spintronics devices despite the fact that no true long-range
magnetic order is possible in one dimension. We study the transverse
and longitudinal fluctuations of magnetic moments at zigzag edges
of graphene from first principles. We find a high value for the
spin wave stiffness $D$~=~2100~meV~\AA$^2$ and a spin-collinear
domain wall creation energy $E_{\rm dw}$~=~114~meV accompanied by
low magnetic anisotropy. Above the crossover temperature
$T_{\rm x}\approx$10~K the spin correlation length $\xi \propto
T^{-1}$ limits the long-range magnetic order to $\sim$1~nm at
300~K while below $T_{\rm x}$ it grows exponentially with decreasing 
temperature. We discuss possible ways of increasing the range of
magnetic order and effects of edge roughness on it.

\end{abstract}

\maketitle

Graphene, a two-dimensional form of carbon, has attracted considerable
attention due to its unique physical properties and potential
technological applications \cite{Geim07,Katsnelson07}. The
possibility of designing graphene-based magnetic nanostructures is
particularly intriguing and has been fuelled by the recent
experimental observations of magnetism in graphitic materials
\cite{Esquinazi03,Ohldag07}. A number of exceptional nanoscale
spintronics devices built around the phenomenon of spin
polarization localized at one-dimensional (1D) zigzag edges of
graphene have been proposed \cite{Son06,Tombros07,Brey07,Wimmer07}. However,
feasibility of such devices is questioned by the fact that no true
long-range magnetic ordering in 1D systems is possible at 
finite temperatures \cite{Mermin66}. Nevertheless, nanometer range spin
correlation lengths in certain 1D systems have been achieved in
practice \cite{Gambardella02}. Establishing the range of magnetic order 
at graphene edges as well as the underlying physical mechanisms is 
thus crucial for practical realization of the proposed spintronics devices.

\begin{figure}
\includegraphics[width=8.5cm]{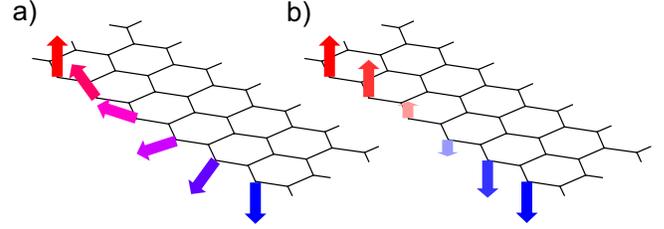}
\caption{\label{fig1} (Color online) Schematic representation of
the transverse (a) and longitudinal (b) low-energy spin
excitation at graphene zigzag edges. The magnetic moments of the
outermost edge atoms are shown by arrows. The direction of
magnetic moments is represented by direction and color of the
arrows while the magnitude is illustrated through the arrow lengths 
and color intensities.}
\end{figure}

In this Letter we study the magnetic correlations at zigzag 
edges of graphene by investigating the transverse and 
longitudinal fluctuations of magnetic moments from first principles. 
While the transverse excitations (spin waves) are characterized by 
the continuous rotation of the electron spin moments along the edge 
(Fig.~\ref{fig1}a), the longitudinal fluctuations affect the spin 
correlation length only if an inversion of magnetic moments resulting 
in appearance of a spin-collinear domain wall \cite{Wakabayashi03}
takes place (Fig.~\ref{fig1}b). The evaluated energies of these 
low-energy excitations mapped onto the classical Heisenberg/Ising 
models allow us to estimate the spin correlation lengths at different
temperatures. Finally, possible ways of increasing the spin correlation 
length and the effects of edge roughness are discussed.

The first-principles calculations of the magnetic excitations are
performed on the density functional theory (DFT) level using the
Perdew-Burke-Ernzerhof exchange-correlation functional
\cite{Perdew96}. A non-collinear spin DFT formalism
\cite{Oda98,Gebauer00} implemented in the \texttt{PWSCF} plane
wave pseudopotential code \cite{PWSCF} in combination with the
ultrasoft pseudopotentials \cite{Vanderbilt90} and a plane wave
kinetic energy cutoff of 25~Ry is used to study spin wave
modes. Much larger supercells are required to obtain converged
results for the spin-collinear domain walls. These calculations are
performed using the standard spin-polarized DFT scheme 
implemented in the \texttt{SIESTA} code \cite{Soler02} together with
a double-$\zeta$ plus polarization basis set, an energy cutoff
of 200~Ry and normconserving pseudopotentials \cite{Troullier91}.
Test calculations performed on limited size systems verify that both
codes provide results in close agreement. The model systems
considered are the hydrogen-terminated periodic one-dimensional
graphene nanoribbons of different widths and supercell lengths
relaxed in their ground state configurations.

\begin{figure}
\includegraphics[width=7cm]{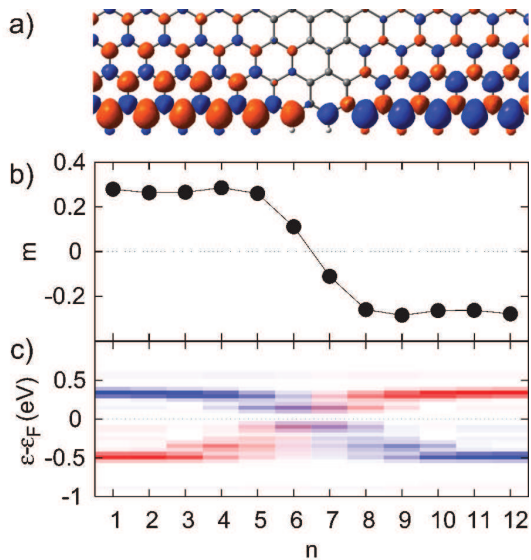}
\caption{\label{fig2} (Color online) (a): Spin density isosurface
plot for the collinear domain wall excitation at a zigzag edge of graphene.
Spin populations $m$ (b) and spin-resolved projected density
of states (c) for the outermost edge atoms. The projected density of
states values for spin-up and spin-down electrons are indicated by the
intensities of red and blue colors, respectively. The edge atoms 
are numbered with $n$. 
}
\end{figure}

The ground state electronic configurations of zigzag graphene nanoribbons 
is characterized by the ferromagnetic arrangement of spins along 
the edges and antiferromagnetic coupling of the spins at the 
opposite edges \cite{Lee05}. To obtain a spin-wave-excited state 
we perform constrained self-consistent calculations with a penalty 
functional term \cite{penalty} added to the total energy expression 
in order to induce small non-collinear deviations of the magnetization 
directions from the spin-collinear ground-state configuration. The total
energy energy difference is mapped onto the quadratic spin-wave
dispersion relation, $E(q)=\kappa q^2$, with
$\kappa=320$~meV~\AA$^2$. At a zigzag edge of graphene the magnetic
moments of the outermost edge atoms $m_{\rm edge}=0.28$~$\mu_{\rm B}$
while the magnetic moments localized on the atoms belonging 
to the $A$ and $B$ sublattices within a single edge unit cell are 
$m_{\rm A}=0.43$~$\mu_{\rm B}$ and $m_{\rm B}=-0.13$~$\mu_{\rm B}$,
respectively. This yields a total magnetic moment of $m=m_{\rm
A}+m_{\rm B}=0.30$~$\mu_{\rm B}$ per unit cell of zigzag
edge. The obtained value of $m$ agrees with the fact that
in zigzag graphene nanoribbons a flat band develops in
one-third of the 1D Brillouin zone ($2\pi/3 \leq | k
a_z | \leq \pi$; $a_z=2.46$~\AA\ is the unit cell length) when
electron-electron interactions are not taken into account
\cite{Nakada96}. The spin-wave stiffness constant $D = 2\kappa /m$
turns out to be 2100 meV~\AA$^2$. Actually, this is a very high
value which is about one order of magnitude higher than the stiffness
constant of bcc iron
\cite{Lynn75,You80}, a three-dimensional solid with much larger
magnetic moment of 2.2~$\mu_{\rm B}$ per atom. Thus, our results confirm
the expectation of higher spin stiffness values in magnetic
materials based on $sp$ elements compared to $d$ element materials
\cite{Edwards06}.

In $sp$-electron itinerant-electron magnets, Stoner-type
longitudinal spin fluctuations may be essential \cite{Edwards06}.
To estimate their characteristic energy we study collinear
domain walls at the graphene zigzag edge. We have performed the
calculations on a large graphene nanoribbon supercells (up to
$\approx$1.8~nm wide and 6~nm long). In order to converge the
self-consistent calculations to the domain wall solution we
provide an appropriate initial magnetizations of edge atoms with
two equidistant domain walls per unit cell for maintaining 
periodicity along the nanoribbon direction. Figure~\ref{fig2}a
illustrates the distribution of the spin density at such a domain
wall located in the center of the edge fragment shown. The spin
populations of the outermost edge atoms (Fig.~\ref{fig2}b) show
that the domain wall is practically localized within two unit
cells (0.5~nm) and the magnetization exhibits weak oscillations
close to the kink. The spin-resolved projected density of states
for the outermost edge atoms (Fig.~\ref{fig2}c) shows an
avoided crossing pattern with band gap diminishing (but not
closing) at the domain wall. From the total energy difference we find
a collinear domain wall creation energy $E_{\rm dw}$~=~114~meV per
edge.

In order to determine the magnetic correlation parameters in the presence
of spin wave fluctuations we recall the nearest-neighbor 1D classical
Heisenberg model
\begin{equation}
    H = -a \sum_{i} \hat{\mathbf s}_i \hat{\mathbf s}_{i+1} - d \sum_{i} \hat{s}^z_i \hat{s}^z_{i+1} - m{\mathbf H}\sum_{i} \hat{\mathbf s}_i,
\end{equation}
where $\hat{\mathbf s}_i$ is the magnetic moment unit vector at site
$i$ and ${\mathbf H}$ is the external magnetic field vector. 
The Heisenberg coupling $a=2\kappa/a_z^2$~=~105~meV
corresponds to the value of $\kappa$ calculated above from 
first principles. The axial anisotropy parameter $d$ is expected 
to be small due to intrinsically weak spin-orbit coupling in graphene
\cite{Huertas-Hernando06,Min06}. We obtain an-order-of-magnitude
estimate for the magnetic anisotropy $d/a=10^{-4}$ using the
spin-orbit coupling strength of $\sim$0.01~meV \cite{Huertas-Hernando06} 
predicted for graphene with weak corrugations 
observed experimentally \cite{Morozov06,Meyer07}. The estimated
$d/a$ agrees with the recent measurements of 2D magnetic
correlations in irradiated graphite \cite{Barzola-Quiquia07} and
with the electron spin resonance $g$-tensor anisotropies 
in molecular graphitic radicals \cite{Stone64,Segal69}. 

\begin{figure}
\includegraphics[width=7cm]{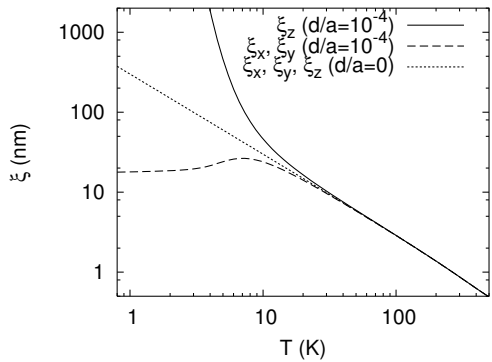}
\caption{\label{fig3} Correlation lengths of magnetization vector 
components orthogonal ($\xi_z$) and parallel ($\xi_x$, $\xi_y$) 
to the graphene plane as a function of temperature $T$ for weakly 
anisotropic ($d/a=10^{-4}$) and isotropic ($d/a=0$) Heisenberg models.
}
\end{figure}

The spin correlation length $\xi^{\alpha}$ ($\alpha=x$, $y$, $z$) 
defines the decay law of the spin correlation
function $\langle \hat{s}^{\alpha}_i \hat{s}^{\alpha}_{i+l}
\rangle = \langle \hat{s}^\alpha_i \hat{s}^\alpha_{i} \rangle
{\rm exp}(-l/\xi^{\alpha})$, i.e.\ the range of magnetic order.
First, we evaluate the zero-field spin correlation length due to 
the transverse spin fluctuations as a function of temperature (see
Fig.~\ref{fig3}) \cite{Joyce67}. Above the crossover temperature
$T_{\rm x} = \sqrt{ad} \approx$ 10~K \cite{Faria79} the small
anisotropy term of the model Hamiltonian has practically no influence
and the system exhibits behavior typical for an isotropic
Heisengerg model \cite{Fisher64} with $\xi^\alpha_{\rm
sw} \approx 300/T$~[nm] and $\langle \hat{s}^\alpha_i
\hat{s}^\alpha_{i} \rangle=$ 1/3. Below $T_{\rm x}$ the anisotropy
term starts playing an important role and the solution exhibits
a characteristic for 1D Ising model exponential divergence of
$\xi^z_{\rm sw} \propto {\rm exp}(\sqrt{8ad}/kT)$ and $\langle
\hat{s}^z_i \hat{s}^z_{i} \rangle=$~1 for $T \rightarrow$~0~K. 
The spin correlation length at zero field in the presence of spin-collinear 
domain walls is the one for 1D Ising model, $\xi^\alpha_{\rm
dw} \approx{\rm exp}(E_{\rm dw}/kT)$. Since $E_{\rm
dw} \gg \sqrt{8ad}$ \cite{Faria79} the overall spin correlation
length $\xi$ in the presence of both transverse and longitudinal
fluctuations, $\xi^{-1}=\xi_{\rm sw}^{-1} + \xi_{\rm
dw}^{-1} \approx \xi_{\rm sw}^{-1}$ is defined predominantly by the
spin wave disorder.

At room temperature ($\sim$300~K) the spin correlation length $\xi=$ 3.7
unit cells ($\sim$1~nm). This result implies that a spintronics
device based on magnetic graphene edges can be operated at room
temperature only if its dimensions do not exceed several spin correlation
lengths, i.e.\ several nanometers. The device dimensions can be scaled
linearly by lowering the operation temperature and below $T_{\rm x}$
this size could be extended beyond the micrometer scale. These estimations
may first look rather disappointing, but nevertheless they are comparable to
one of the most appealing example of 1D magnetism: monoatomic Co chains
on Pt substrate characterized by a ferromagnetic order range of
$\approx$4~nm at 45~K \cite{Gambardella02}. In this $d$-element
system ferromagnetic order stems mainly from the anomalously
high magnetic anisotropy which is absent in graphene nanostructures.
However, the lack of anisotropy is partially compensated by the
high spin stiffness which results in considerable spin correlation
lengths even in the isotropic regime above $T_{\rm x}$. While the spin
stiffness constant can hardly be increased we suggest several
ways of increasing the magnetic anisotropy (and thus $T_{\rm x}$)
by strengthening the spin-orbit coupling by increasing curvature,
applying external electric field or coupling graphene to a substrate
\cite{Huertas-Hernando06}. Alternatively, the magnetic anisotropies
can be increased by chemical functionalization of graphene edges
with heavy element functional groups (e.g. iodine) coupled to the
spin-polarized edges states via the exchange polarization 
\cite{Goodings61,Yazyev07b}. Augmenting the crossover temperature 
above 300~K would result in a significant increase of $\xi^z$ to the 
length scales of the present-day semiconductor technology.

Thus, the graphene edges at finite temperatures are not actually 
ferromagnetic but superparamagnetic ones. For the isotropic Heisenberg 
model the enhancement factor for the susceptibility in comparison 
with one of noninteracting spins reads \cite{Fisher64}
\begin{equation}
\frac{\chi}{\chi_0} = \frac{1+u}{1-u} \approx \frac{2a}{T}
\end{equation}
where $u = {\rm coth}(a/T) - T/a$ and the approximation being 
valid at $a \gg T$. At room temperature the susceptibility enhancement 
factor $\chi/\chi_0\approx$ 8.

\begin{figure}
\includegraphics[width=8.5cm]{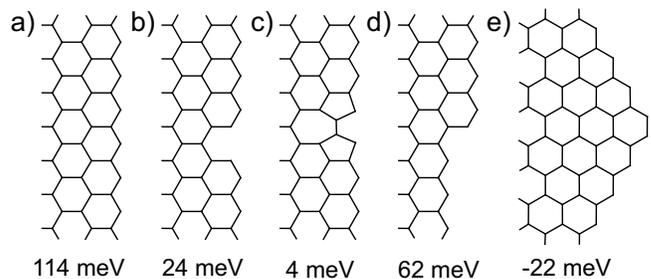}
\caption{\label{fig4} Ideal zigzag edge of graphene (a) and various
types of edge defects: missing or rehybridized edge atom (b),
Stone-Wales defect (c), edge step (d), and 120$^\circ$ edge turn.
The domain wall creation energies at these structures are shown.
}
\end{figure}

Although we found a relatively high value of $E_{\rm dw}$, the
localized domain walls may become energetically more favorable at
edge defects, and therefore we discuss creation of localized domain walls
at different types of topological imperfections at zigzag graphene
edge classified as shown in Fig.~\ref{fig4}. The simplest case of
edge roughness is a boundary atom missing from the
$\pi$-conjugation network (Fig.~\ref{fig4}b). Such $sp^2$-vacancy
formation may result from the rehybridization of an outermost atom
into the $sp^3$ state due to chemical modification or because of
the creation of a true vacancy. The domain wall creation energy at
an $sp^3$-hybridized atom is found to be 24~meV, i.e.\ factor of 5 
smaller than $E_{\rm dw}=$ 114~meV for the ideal
zigzag edge. Such decrease will have a dramatic effect on the
long-range magnetic order at room temperature since $E_{\rm dw}$
is lowered to $kT$ ($\approx$25~meV at 300~K). An even more dramatic
decrease to 4~meV is observed at the Stone-Wales defect
(Fig.~\ref{fig4}c), a topological structure obtained by the
90$^\circ$-rotation of a single C$-$C bond which locally breaks
the bipartite lattice symmetry. The presence of an edge step
(Fig.~\ref{fig4}d) has a less severe effect and reduces $E_{\rm dw}$
to 62~meV. A completely different situation is observed for a
120$^\circ$-turn of the zigzag edge (Fig.~\ref{fig4}e).
The antiferromagnetic arrangement of spins at the edge segments
separated by the 120$^\circ$-turn is by 22~meV more stable than the
ferromagnetic arrangement. This is due to the change of bipartite sublattice
to which belong the outermost edge atoms and due to the antiferromagnetic 
coupling between the magnetic moments in different sublattices 
\cite{Lieb89,Yazyev07}. Similar behavior 
has recently been pointed out for the edges of hexagonal 
graphene nanoislands \cite{Fernandez-Rossier07}. 
Domain walls are thus naturally pinned to such turns, although 
the energy difference is close to $kT$ at room temperature. 
A ``spin-inverter'' device design based on such a 120$^\circ$-turn
topology can be anticipated. Simple chemical modifications which do 
not perturb the $\pi$ conjugation network at graphene edges
show almost no effect on $E_{\rm dw}$. For an ideal zigzag edge
terminated with electronegative fluorine atoms we find $E_{\rm
dw}=$ 117~meV very close to the value for the hydrogen-terminated edge 
(114~meV).

To conclude, we have studied from first principles the energetics of
transverse and longitudinal spin fluctuations at the one-dimensional
magnetic zigzag edge of graphene. The transverse fluctuations
characterized by the high spin stiffness constant are the main
limiting factor of the spin correlation length which is found to be $\sim$1~nm
at room temperature. For the temperatures above $\sim$10~K the spin
correlation length is inversely proportional to the temperature due
to the low magnetic anisotropy of the system. Below the crossover
temperature the spin correlation length grows exponentially with
decreasing temperature. We propose several approaches for 
extending the range of magnetic order by increasing the magnetic 
anisotropy in this carbon-based system and discuss the effect 
of edge roughness on the spin correlation length.

We wish to thank L.~Helm for his critical reading of the manuscript. 
M.~I.~K. acknowledges financial support from FOM (the Netherlands).
The computational resources were provided by the Swiss National 
Supercomputing Center (CSCS). 


\begin{thebibliography}{39}
\bibitem{penalty}  
The penalty term $E^p=\lambda\sum_i([\hat{m}_{ix}-{\rm cos}(\theta_i)
{\rm sin}(\phi_i)]^2+[\hat{m}_{iy}-{\rm sin}(\theta_i) {\rm sin}(\phi_i)]^2+
[\hat{m}_{iz}-{\rm cos}(\phi_i)]^2)$
used in this work induces a precession-like arrangement of spins 
along the graphene edge by constraining the magnetic moment 
direction $\hat{{\mathbf m}}_i$ of each carbon atom $i$. 
The deviation angle $\phi_i=\Phi$ if atom $i$ belongs to 
sublattice $A$ and $\phi_i=\Phi+\pi$ otherwise. The angle $\theta_i$ 
varies with the fractional coordinate $a_i$ of the 1D supercell 
as $\theta_i=2\pi a_i$. The resulting magnitude of spin 
stiffness $D$ is stable against the choice of 
$\lambda$ and $\Phi$ equal here to 10~Ry and $\pi/36$, respectively.
\end{thebibliography}

\end{document}